\documentclass[aoas]{imsart}

\RequirePackage{amsthm,amsmath,amsfonts,amssymb}
\RequirePackage[authoryear]{natbib}
\RequirePackage[colorlinks,citecolor=blue,urlcolor=blue]{hyperref}
\RequirePackage{graphicx}

\usepackage{mathtools}
\usepackage{algorithm}
\usepackage{algpseudocode}
\usepackage{eqnarray}
\usepackage{enumitem}

\startlocaldefs

\newcommand{\ba}{\begin{eqnarray*}}
\newcommand{\ea}{\end{eqnarray*}}
\newcommand{\be}{\begin{eqnarray}}
\newcommand{\ee}{\end{eqnarray}}

\theoremstyle{plain}

\newtheorem{theorem}{Theorem}[section]

\theoremstyle{remark}
\newtheorem{definition}[theorem]{Definition}
\newtheorem{corollary}[theorem]{Corollary}

\endlocaldefs

\begin{document}
\begin{frontmatter}

\title{Multiple Hypothesis Testing To Estimate The Number of Communities in Sparse Stochastic Block Models}
\runtitle{Multiple Testing Approach for sparse SBMs}

\begin{aug}
\author[A]{\fnms{Chetkar} \snm{Jha}\ead[label=e1]{Chetkar.Jha@pennmedicine.upenn.edu}},
\author[A]{\fnms{Mingyao} \snm{Li}\ead[label=e2]{mingyao@pennmedicine.upenn.edu}},
\and
\author[A]{\fnms{Ian} \snm{Barnett}\ead[label=e3]{ibarnett@pennmedicine.upenn.edu}}
\address[A]{Department of Biostatistics, Epidemiology and Informatics, University of Pennsylvania, USA
\printead{e1,e2,e3}}
\end{aug}

%


\begin{abstract}
Network-based clustering methods frequently require the number of communities to be specified
\emph{a priori}. Moreover, most of the existing methods for estimating the number of communities
assume the number of communities to be fixed and not scale with the network size $n$. The few methods 
that assume the number of communities to increase with the network size $n$ are only valid when the 
average degree $d$ of a network grows at least as fast as $O(n)$ (i.e., the dense case) or 
lies within a narrow range. This presents a challenge in clustering large-scale network data, particularly
when the average degree $d$ of a network grows slower than the rate of $O(n)$ (i.e., the sparse case).
To address this problem, we proposed a new sequential procedure utilizing multiple hypothesis tests and the 
spectral properties of Erd\"{o}s R\'{e}nyi graphs for estimating 
the number of communities in sparse stochastic block models (SBMs). 
We prove the consistency of our method for sparse SBMs for a broad range of the sparsity parameter. 
As a consequence, we discover that our method can estimate the number of communities $K^{(n)}_{\star}$ 
with $K^{(n)}_{\star}$ increasing at the rate as 
high as $O(n^{(1 -  3\gamma)/(4 -  3\gamma)})$, where $d = O(n^{1 - \gamma})$. Moreover, we show that our method can be adapted as a stopping rule in estimating the number of communities in binary tree stochastic block models.
We benchmark the performance of our method against other
competing methods on six reference single-cell RNA sequencing datasets.
Finally, we demonstrate the usefulness of our method through numerical simulations and by using it for 
clustering real single-cell RNA-sequencing datasets.
\end{abstract}


\end{frontmatter}

\section{Introduction}
\subsection{Motivation}
Network clustering or community detection methods have wide applicability 
in many areas of science. In particular, many community detection
based methods are being used for clustering single-cell RNA sequencing (scRNA-seq)
datasets, see \cite{blondel2008}, \cite{Lancichinetti2009}, \cite{satija2015}, \cite{ding2016}, 
and \cite{kiselev2019} for a review. 
Single-cell technology collects sequencing data at the cellular level 
providing a higher resolution of the cellular differences (\cite{eberwine2014}). Clustering of cells, 
based on scRNA-seq datasets, into a fewer number of cell clusters or communities can potentially
inform us about each cluster's functional role and biological relevance.

Moreover, in recent years, we are witnessing rapid advancement in the single-cell technology
resulting in larger collections of scRNA-seq datasets at higher resolutions.
As more and more cellular level data is collected, one would expect the number of communities of scRNA-seq 
datasets to potentially scale with the network size $n$.
This is empirically supported by Figure 5 of \cite{valentine2020}, where they showed that 
the number of estimated cell clusters tend to increase with the total number of cells in a large number
of studies. This problem of estimating a large number of communities is further amplified in the presence of
sparsity, which are an ever-present feature of scRNA-seq datasets (\cite{sparseScRNA}).
However, most of the existing methods for estimating the number of communities in sparse
networks do not allow for the number of communities to increase with the network size $n$.
Motivated by this problem, we proposed a new method for estimating the number
of communities in stochastic block models.

\subsection{Background}
The stochastic block model (SBM), proposed by \cite{holland83},
is a popular model for network data with the block or community structure.
It assigns nodes into different communities and the edge probability between any pair of
nodes is determined by their respective communities.
Recently, many extensions and variants of the SBM were proposed.
For instance, \cite{airoldi2008} proposed the mixed membership stochastic block model
allowing nodes to belong to multiple networks.
\cite{karrer2011} proposed the degree-corrected stochastic
block model (DCSBM) relaxing the assumption of homogeneity of nodes.
\cite{li2020} proposed a binary tree stochastic block model (BTSBM) allowing homogeneous blocks
of SBMs to be arranged in a hierarchical network. 

Currently, there are many existing methods for estimating the community structure 
of SBMs including modularity maximization 
(\cite{newman04}) approaches, Louvain modularity algorithm (\cite{blondel2008}),
likelihood-based approaches (\cite{bickel09}; \cite{zhao12}; \cite{choi2012};\cite{amini2013}), 
spectral clustering methods (\cite{rohe11}; \cite{lei15};\cite{joseph2016}) among others see \cite{zhao17}
for review. Several existing methods such as \cite{newman04}; \cite{bickel09}; \cite{zhao12}; \cite{qin_dcsbm2013};
\cite{amini2013} have also been shown to be consistent for both sparse and dense DCSBMs.
However, all of these existing community detection methods require the true number 
of communities \emph{a priori} for estimating the community structure.

Fortunately, there are several existing methods for estimating the number of communities. 
These methods can be broadly classified into three categories: i) Likelihood-based methods, 
ii) Cross-validation-based methods, and ii) Spectral methods. The likelihood-based methods use the 
likelihood function or approximate pseudo-likelihood function for selecting the best model and thereby 
estimating the number of communities. In particular, \cite{wang17} and \cite{ma2019} proposed a 
likelihood-based model selection (LRBIC) approach and a pseudo-likelihood-based (PL) approach, respectively, for 
estimating the number of communities in both dense and sparse DCSBMs. The cross-validation-based approaches 
use network resampling strategies to generate multiple copies of the network and subsequently use the 
cross-validation method to select the optimum number of communities.
Specifically, \cite{li2016} and \cite{chen2018} proposed an edge cross-validation (ECV) approach 
and a network cross-validation (NCV) approach, respectively,
to estimate the number of communities for both dense and sparse DCSBMs.
The spectral methods utilize the spectral properties of appropriately modified adjacency matrices for estimating
the number of communities. In particular, \cite{lei16} proposed a Goodness-of-fit (GoF) approach utilizing the spectral properties
of the generalized Wigner matrices, whereas \cite{can2015} proposed BHMC and NB approaches utilizing the spectral properties
of the Bethe Hessian matrix and the Non-backtracking matrix, respectively.

In our view, the spectral methods have several distinct advantages over the non-spectral methods. 
In particular, the spectral methods 
allow for the number of communities to increase with the
network size $n$, whereas the non-spectral methods assume the number of communities to be fixed.
Moreover, the spectral methods tend to be robust to the likelihood-based assumptions, and are
also computationally efficient for large networks. The latter is true because spectral methods only require computing
few eigenvalues. The main drawback of the existing spectral methods is that they only provide theoretical guarantees
(such as consistency) of their results when the average degree $d$ of the network grows at least with the 
rate of $O(n)$ (i.e., dense case) or lies within a specific range. For instance, in Erd\"{o}s R\'{e}nyi graphs, 
the BHMC and NB approaches of \cite{can2015} are only valid when the average degree $d$ of the 
network satisfies $d < n^{2/13}$. 
Moreover, the GoF approach of \cite{lei16} is not even applicable for sparse Erd\"{o}s R\'{e}nyi graphs.

Parallel to the above developments, several authors such as \cite{clauset2008}, \cite{peel2015},
\cite{li2020} proposed hierarchical networks for modeling a large number of communities.
In hierarchical networks, the number of communities increases at the rate of the order of
the exponent of the depth (i.e., resolution) of the network. 
Recently, \cite{li2020} proposed a special type of hierarchical network called binary tree stochastic
block model (BTSBM), where the hierarchical network has a binary tree structure with Erd\"{o}s R\'{e}nyi
graphs as its leaves. 
For retrieving the community structure in BTSBMs, \cite{li2020} proposed recursively bipartitioning
the network until a stopping criterion is reached, see \cite{li2020}.  
Popular methods for bipartitioning a network into sub-networks 
utilize either the sign of the second eigenvalues
of the adjacency matrix or use spectral methods such as
regularized spectral clustering method, see
 \cite{balakrishnan2011}, \cite{gao2017}, \cite{amini2013}, etc.
On the other hand, the stopping criterion is used to determine whether further bipartitioning is
required. \cite{li2020} showed that the NB method of \cite{can2015} can be used as a stopping rule.
As we discuss, the above hierarchical methods do not fare better in comparison to our
proposed method for estimating a large number of communities in SBMs.

Motivated by the challenges in clustering large and sparse single-cell RNA sequencing datasets, we proposed a
new spectral approach for estimating the number of communities in sparse and dense SBMs. 
Our approach is based on the observation that a SBM consisting of $K$ blocks is equivalent
to stating that a SBM consists of $K$ distinct Erd\"{o}s R\'{e}nyi blocks. To avoid any ambiguity
in identifying Erd\"{o}s R\'{e}nyi blocks in a SBM, we require that the edge probability with which
edges are formed within Erd\"{o}s R\'{e}nyi blocks be strictly greater than
the edge probability with which edges are formed between a pair of different Erd\"{o}s R\'{e}nyi blocks.
Then, it immediately follows that estimating the number of blocks in a SBM is equivalent to estimating 
the number of Erd\"{o}s R\'{e}nyi blocks within the SBM.
We use this idea to estimate the number of blocks in a SBM. 
In particular for testing whether the SBM has $K_0$ Erd\"{o}s R\'{e}nyi blocks (i.e., $H_0$), 
we proposed a multiple hypothesis test simultaneously testing whether all $K_0$ distinct blocks 
within the SBM are Erd\"{o}s R\'{e}nyi. Subsequently, we use this test sequentially
at every value of $K_0$ to determine whether a SBM has $K_0$ 
Erd\"{o}s R\'{e}nyi blocks (i.e., $H_0$), where $K_0$ is incremented by one (starting with $K_0=1$) until
the test fails to reject $H_0$. 
As we discuss later, the above multiple hypothesis test utilizes \cite{lee2018} and their result
on the second largest eigenvalue of an Erd\"{o}s R\'{e}nyi graph and is adapted for our use.

Our main contributions are as follows: i) We proposed a new sequential testing procedure (SMT)
for estimating a large number of communities in sparse SBMs. ii) We proved that our estimator is consistent
for estimating the  true number of communities $K_{\star}^{(n)}$ while allowing for $K^{(n)}_{\star}$
to increase  at a rate of $O( n^{(1 - 3\gamma)/(4 - 3\gamma)})$, where $\rho_n= O(n^{-\gamma})$ is the sparsity
parameter of the network and $\gamma$ is a constant. iii) Moreover, we showed that our method can be used as a stopping rule
for estimating the number of communities in BTSBMs.
Although we have applied our approach for clustering scRNA-seq datasets, our
approach is general and can be used for other datasets.
The rest of the paper is organized as follows. Section \ref{s.pre} gives the necessary
model and notational definitions and describes the preliminary set up for our analysis.
Section \ref{s.main} establishes the consistency result for SMT for sparse SBMs, and extends
our method for hierarchical networks.
Section \ref{s.num} compares the performance of SMT against competing methods for
small and large sparse network datasets. Moreover, Section \ref{s.num} compares the hierarchical
version against the competing methods on hierarchical networks.
Section \ref{s.data} benchmarks the performance of SMT by comparing its results on six reference
single-cell datasets. Section \ref{s.data} also uses SMT and the hierarchical variant of SMT for clustering
of real scRNA-seq datasets. Section \ref{s.disc} concludes the paper with a discussion.

\section{Preliminaries}\label{s.pre}

\subsection{Stochastic Block Model}
A SBM for a network of $n$ nodes with $K^{(n)}_{\star}$ blocks
is parametrized by a block membership vector $g = \{1, \cdots, K^{(n)}_{\star} \}^n$ and
a symmetric block-wise edge probability matrix $G_{\rho_n}$, where $\rho_n = O(n^{-\gamma})$
is the sparsity parameter and $\gamma$ is a constant. The sparsity parameter $\rho_n$ dictates the average degree of
the network of size $n$. With some foresight, $\delta_0$ be the maximum difference between within block 
probabilities and between block probabilities over $n$ nodes, i.e.,
\be\label{delta0}
\delta_0 = \max_{1 \le i \le K^{(n)}_{\star}}\{ \max_{1 \le j \le K^{(n)}_{\star}}(G_{\rho_n}(i,i) -  G_{\rho_n}(i,j)) \}.
\ee

A SBM assumes that the probability of an edge between any pair of nodes $i,j$ is given by the 
edge probability between their respective blocks $g_i$ and $g_j$. Thus, we have the following relation between
the node-wise edge probability matrix $P$ and block-wise edge probability matrix $G_{\rho_n}$. 
\ba
P( (i, j) \mid g_i = k, g_j = k') = G_{\rho_n}(k,k'), 1 \le i,j \le n, 1 \le k,k' \le K^{(n)}_{\star}.
\ea

Let $A =\{0, 1\}^{n \times n}$ be the observed symmetric adjacency matrix with no self-loops, (i.e., $A_{ii}=0$), for $1 \le i \le n$.
Letting every edge given $(g, G_{\rho_n})$ (up to a symmetric constraint $A_{ij} = A_{ji}$) be an independent Bernoulli random variable, the probability mass function of the adjacency matrix $A$, given $(g, G_{\rho_n})$ is:
\ba
P_{g, G_{\rho_n}}(A)= \Pi_{1 \le i <j \le n} ( G_{\rho_n}(g_i, g_j))^{A_{ij}}(1 - G_{\rho_n}(g_i,g_j))^{1- A_{ij}}.
\ea

\subsection{Main Idea}
Our approach is based on the observation that a SBM with $K^{(n)}_{\star}$ blocks consists of $K^{(n)}_{\star}$ distinct Erd\"{o}s R\'{e}nyi blocks. To avoid any ambiguity in identifying Erd\"{o}s R\'{e}nyi blocks in a SBM, we require that the edge probability with which edges are formed within Erd\"{o}s R\'{e}nyi blocks be strictly greater than the edge probabilities between nodes
from different blocks. Then, it immediately follows that estimating the number of blocks of the SBM is equivalent to estimating the number of distinct  Erd\"{o}s R\'{e}nyi blocks within the SBM. 
We use this insight to propose a sequential multiple testing (SMT) approach for estimating the number of blocks of a SBM.

Let $N_i$ be the total number of nodes belonging to $i^{th}$ block, i.e. $N_i = \{ \sum_{j=1}^n g(j)= i \}$.
Let $\{A^{(i)}_{\star} =\{0,1\}^{N_i \times  N_i}\}_{i=1, \cdots, K^{(n)}_{\star}}$ denote $K^{(n)}_{\star}$ adjacency matrices corresponding to $K^{(n)}_{\star}$  Erd\"{o}s R\'{e}nyi blocks within the network. Let $\{ p_i \}_{i=1}^K$ denote the common nodewise probability with which the edges are formed within the $K^{(n)}_{\star}$ Erd\"{o}s R\'{e}nyi blocks. Then, we define scaled adjacency matrices $\{ M^{(i)} \}_{i=1}^{K^{(n)}_{\star}}$ as follows.

\be\label{scaleER}
M^{(i)}(u,v) =
\begin{cases}  \frac{1}{\sqrt{N_i p_i (1- p_i)} } &, \text{ if } A^{(i)}_{\star}(u,v) =1, u,v = 1, \cdots, N_i,\\
    0 &, \text{ if } A^{(i)}_{\star}(u,v) =0, u,v = 1, \cdots, N_i,
\end{cases}
\ee
where $i =1, \cdots, K^{(n)}_{\star}$, $K^{(n)}_{\star}$ denotes the total number of  Erd\"{o}s R\'{e}nyi blocks. 

\subsubsection{Second Largest Eigenvalue of Erd\"{o}s R\'{e}nyi Graphs}\label{sec.consis}
Our approach uses the limiting distribution of the second eigenvalue of scaled
adjacency matrices generated from Erd\"{o}s R\'{e}nyi graphs.
To this end, we collect an important result concerning
the second largest eigenvalue of such scaled adjacency matrices.

\begin{theorem}[\cite{lee2018}]\label{er.thm}
Let $A^{(i)}_{\star}$ be the adjacency matrix generated from Erd\"{o}s R\'{e}nyi graph with $N_i$ nodes
with $p_i$ denoting the common node-wise probability within the Erd\"{o}s R\'{e}nyi graph. Assume that for arbitrarily
small $\epsilon > 0$,  $N_i$ and $p_i$ satisfy $ N_i p_i \ge N_i^{1/3 + \epsilon}$. Define $M^{(i)}$ as the scaled adjacency matrix in (\ref{scaleER}) and $\mu_i = N_i p_i$. Then, the second largest eigenvalue of $M^{(i)}$ obeys the \emph{Tracy-Widom} distribution with a deterministic shift $\mu_i$, i.e.,
\be
N_i^{2/3} (\lambda_2(M^{(i)}) - 2 - 1/\mu_i) \xrightarrow{ } TW_1(\cdot), 
\ee
where $TW_1(\cdot)$ is the \emph{Tracy-Widom} distribution with Dyson parameter one.
\end{theorem}

The above theorem characterizes Erd\"{o}s R\'{e}nyi graphs using the limiting distribution
of the second largest eigenvalue of the scaled adjacency matrix. In particular, the above theorem
is valid  when the average degree $d$ of the network satisfies $d \ge \min_{i=1, \cdots, K}O(N_i^{1/3})$, where $N_i$
is the total number of nodes in the $i^{th}$ Erd\"{o}s R\'{e}nyi block.
Unfortunately, Theorem \ref{er.thm} is given in terms of an unknown scaled adjacency matrix $M^{(i)}$ which is a function
of an unknown parameter $G_{\rho_n}(i,i)$. We give an estimable version of Theorem \ref{er.thm} in Section \ref{s.main}.
It is worth noting that we can estimate $G_{\rho_n}$ provided we have a consistent
estimator of the community membership vector $g$ of the original observed adjacency matrix. Definition \ref{cons.det} gives a consistent estimator of the community membership vector $g$.
In this regard, we know that for fixed $K_{\star}$, several methods can recover true
communities, such as the profile likelihood method \citep{bickel09} and the spectral clustering method
\citep{lei2014}. For $K^{(n)}_{\star}$ increasing with $n$, some methods can recover true communities for
some special cases such as planted partition models, see \cite{chaudhuri2012}, \cite{amini2018}.
Moreover, for sparse SBMs, community detection methods
such as those of \cite{newman04}, \cite{bickel09}, \cite{zhao12}, and \cite{amini2013} 
can also recover true communities.

\begin{definition}\normalfont\label{cons.det}[Consistency of Community Detection]
A sequence of stochastic block models, indexed by $\{(g^{(n)}, G^{(n)}_{\rho_n}), n \ge 1 \}$
with $K^{(n)}_{\star}$ communities, is said to have a consistent community membership estimator
$\hat{g}(A, K^{(n)}_{\star})$ if:
\ba
\lim_{n\to \infty} P(\hat{g} =g^{(n)} \mid A \sim (g^{(n)}, G^{(n)}_{\rho_n})) \to 1,
\ea
where $g^{(n)}$ and $G^{(n)}_{\rho_n}$ denote a sequence of community membership vector and blockwise edge
probability matrix increasing with the network size $n$, respectively.
\end{definition}

\subsection{Sequential Test For Estimating Number of Communities}
Based on the ideas discussed in the previous subsection, we give a new procedure for estimating a number of communities.
Let $\{\hat{p}_i\}_{i=1}^K$ be the common node-wise probability for $K$
Erd\"{o}s R\'{e}nyi blocks. Then, we define the estimated scaled adjacency matrices corresponding to $M^{(i)}$ 
as follows

\be\label{estM}
\hat{M}^{(i)}(u,v) =
\begin{cases}  \frac{1}{\sqrt{N_i \hat{p}_i (1- \hat{p}_i)}} &, \text{ if } A^{(i)}_{\star}(u,v) =1, u,v = 1, \cdots, N_i,\\
    0  &, \text{ if } A^{(i)}_{\star}(u,v) =0, u,v = 1, \cdots, \{ \sum_{j=1}^n \hat{g}(j)= i \},
\end{cases} 
\ee
where $i =1, \cdots, K$, $K$ denotes the total number of  Erd\"{o}s R\'{e}nyi block, $\hat{g}$ denotes the community membership vector, and $N_i$ denotes the number of nodes in the $i^{th}$  Erd\"{o}s R\'{e}nyi block.

We test whether all  $K$ blocks are Erd\"{o}s R\'{e}nyi sequentially in A to estimate the number of communities
in a SBM. Our sequential procedure is given in Algorithm \ref{alg:1}.

\begin{algorithm}
\caption{Sequential Multiple Test (SMT) Procedure for SBM}
\label{alg:1}
\begin{algorithmic}[1]
\State\label{step1}Initialize $\hat{K} =1$.
\State\label{step2}Use $\hat{K}$ to obtain an estimate of the community membership vector, $\hat{g}$.
\State Compute the estimated block-wise probability matrix $\hat{G}_{\rho_n}$ using $\hat{g}$, where $\hat{g}$ is computed under the assumption that $K=\hat{K}$.
\State Using $\hat{g}$, define adjacency matrices $A^{(i)}_{\star}$ for every  Erd\"{o}s R\'{e}nyi block, where $i = 1, \cdots, K$.
\State Using (\ref{estM}), define estimated scaled adjacency matrices $M^{(i)}$.
\State Compute $\hat{G}$ as follows
\be\label{est.mod}
\hat{G}_{\rho_n}(k,k') = \frac{\sum_{(s,t) : \hat{g}(s) =k, \hat{g}(t)=k'} A(s,t)}{\sum_{(s,t) : \hat{g}(s) =k, \hat{g}(t)=k'}1}, 1\le k, k' \le K.
\ee
\State  Compute the test statistic $T_{n,K}$ by estimating $\hat{G}_{\rho_n}$ in (\ref{est.mod}) as
\be\label{t.stat}
T_{n, K} = \max_{i=1, \cdots, \hat{K}} N^{\frac{2}{3}}_i( \lambda_2( \hat{M}^{(i)}) - 2 - \frac{1}{\mu_i}) ,
\ee
where $\mu_i =N_i*G_{\rho_n}(i,i)$ and $N_i$ is the number of nodes in the $i^{th}$ Erd\"{o}s R\'{e}nyi block, i.e. $\hat{g}$ satisfies $N_i =\sum_{j=1}^n 1_{\hat{g}(j) =i}$.
\State For a specified nominal significance level $\alpha$, conduct the multiple comparison test
\be
&&\label{ht}  H_0: \text{ All $K$ blocks  are Erd\"{o}s R\'{e}nyi}.\\
&& \label{ht1} H_1: \text{$\exists$ at least one block that is not Erd\"{o}s R\'{e}nyi }.
\ee
\State Accept $K=\hat{K}$, when $T_{n, \hat{K}} \le TW_1(1 - \alpha)$, and stop.
\State\label{step-last} If the test is rejected at the previous step, then increment $\hat{K} = \hat{K} + 1$ and go to Step \ref{step2}.
\end{algorithmic}
\end{algorithm}

\section{Main Results}\label{s.main}

\subsection{Asymptotic Null Distribution}
For obtaining an estimable version of Theorem \ref{er.thm} (given in terms of estimated second eigenvalue), we make use of Weyl's inequality to bound the error incurred because of the estimation. Additionally, we assume the following.

\begin{enumerate}[leftmargin=0pt,itemindent=2em, label =A.\arabic*]
\item\label{a1}(Balancedness) Assume that all the communities of a SBM are balanced, i.e., every community has a similar number of nodes belonging to it in the following sense
\ba
O(N_1) = \cdots = O(N_i)= \cdots = O(N_{K_{\star}}) = O(\frac{n}{K^{(n)}_{\star}}),
\ea
where $N_i$ denotes the total number of nodes belonging to the $i^{th}$ block of the SBM, $K^{(n)}_{\star}$ is the total number of communities, and $n$ is the network size.
\end{enumerate}

\begin{theorem}\normalfont \label{teststat.dist}[Asymptotic Null Distribution]. Let A be an adjacency matrix generated
from a SBM ($g$, $G_{\rho_n}$) with $K^{(n)}_{\star}$ satisfying assumption \ref{a1}, where $\rho_n = O(n^{-\gamma})$. Moreover, assume 
that $\hat{g}$ is a consistent estimate of $g$. Let $\{ \hat{M}^{(i)} \}_{i=1}^{K^{(n)}_{\star}}$ denote $K^{(n)}_{\star}$ estimated scaled adjacency matrices corresponding to the adjacency matrices $\{ A^{(i)}_{\star} \}_{i=1}^{K^{(n)}_{\star}}$, where each $A^{(i)}$ is the adjacency matrix corresponding to the $i^{th}$ Erd\"{o}s R\'{e}nyi block and $i=1, \cdots, K^{(n)}_{\star}$.
Suppose $\min_{i}\{ \hat{G}_{\rho_n}(i,i)  \}\ge (n/K_{\star}^{(n)})^{-2/3}$, $K^{(n)}_{\star} \le n^{\frac{1 - 3\gamma }{4 - 3\gamma}}$, then the second largest eigenvalue of the $\hat{M}^{(i)}$ converges to the \emph{Tracy-Widom} distribution with a deterministic shift $\hat{\mu}_i$, i.e.,
\be\label{main.thm}
(n/K_{\star}^{(n)})^{\frac{2}{3}} \left(\lambda_2(\hat{M}^{(i)}) -2 - \frac{1}{\hat{\mu}_i}\right)\xrightarrow{D} TW_1, \forall i =1, \cdots, K_{\star}, 
\ee
where $\hat{\mu}_i = (n/ K_{\star}^{(n)}) \hat{G}_{\rho_n}(i, i)$ is the estimated deterministic shift and $\xrightarrow{D}$ denotes convergence in distribution.

\end{theorem}

\begin{proof}
The proof is given in the Supplement.
\end{proof}


Like Theorem \ref{er.thm}, Theorem \ref{teststat.dist} assumes that for the true number of communities $K^{(n)}_{\star}$ (i.e., under the null $H_0: K = K_{\star}^{(n)}$) we can consistently recover the true community structure, which is a common assumption in the literature, e.g., see \cite{lei16}, \cite{wang17}, etc.
Theorem \ref{teststat.dist} shows that the centered and scaled second eigenvalues of estimated scaled adjacency matrices
corresponding to the $K_{\star}^{(n)}$ Erd\"{o}s R\'{e}nyi blocks  converge to the \emph{Tracy-Widom} distribution. 
The proof of the Theorem \ref{teststat.dist} follows from minimizing the total sum of committed errors in estimating $K_{\star}^{(n)}$ scaled adjacency matrices. This automatically gives us the condition on the maximum rate at which $K_{\star}^{(n)}$ increases with
$n$. Meanwhile, the condition on $\hat{G}_{\rho_n}$ follows from Theorem \ref{er.thm}.
The downside of our approach is that it only covers the range for which $\min_{i}( \hat{G}_{\rho_n}(i,i) )\ge (n/K_{\star}^{(n)})^{-2/3}$ and does not cover the ultra sparse case between $\{ O(log(n/K_{\star}^{(n)}))$, $ O((n/K_{\star}^{(n)})^{-2/3}) \}$. 

\subsection{Asymptotic Power}
Recall that the estimate of $K^{(n)}_{\star}$ is given as
\be\label{hatk}
\hat{K} = \inf_{k} \{ k : k \in \mathcal{N} : T_{n, k} \le t_{1 - \alpha} \},
\ee
where $T_{n, k}$ is given in (\ref{t.stat}),$t_{1-\alpha}$ is \emph{Tracy-Widom} distribution quantile at $(1- \alpha)$, and
$\mathcal{N}$ is the set of natural numbers.

Recall that $\hat{K}$ in (\ref{hatk}) is a sequential estimate which is incremented by one starting with
$\hat{K} =1$. In the previous subsection, we showed that our estimate $\hat{K}$ under the null  is consistent, i.e.,  $P_{H_0}(\hat{K} = K^{(n)}_{\star}) \to 1.$
Therefore to prove the consistency of $\hat{K}$ in (\ref{hatk}), it is sufficient to show that the power of the 
test when $K < K^{(n)}_{\star}$ (i.e., when SBM is under-fitted) asymptotically goes to one. Unlike Theorem \ref{teststat.dist}, for proving the asymptotic power  we do not make any assumption about the consistent recovery of communities
and only require assumption \ref{a1} about balanced block sizes.

\begin{theorem}\label{power}
The power of the hypothesis test in testing $H_0$  vs $H_1$ (\ref{ht})-(\ref{ht1}) when the true number of blocks
$K < K_{\star}^{(n)}$ and the true within-block probabilities satisfy $\max_{1 \le i \le K_{\star}} G_{\rho_{n, K_{\star}}}(i,i) \in (0, 0.5]\cup (0.5 + \max_{1 \le j \le n}\delta_{0}, 1]$, asymptotically goes to one, i.e.
\be
P_{H_1} (T_{n,K} \ge t_{1 - \alpha}) \to 1,
\ee
where $T_{n, K}$ is the test statistics in (\ref{t.stat}) and $\delta_{0}$ is the maximum difference between
within block probabilities and between block probabilities over $n$ nodes defined in (\ref{delta0}).
\end{theorem}

\begin{proof}
The proof is given in the Supplement.
\end{proof}

The blind spot of the test statistic in (\ref{t.stat}) occurs when the within-block edge probability is greater than $1/2$ and less than $1/2 + \delta_0$, where the asymptotic power does not go to one and our test procedure will not be consistent. 
This is expected because in this case, the between-block edge probabilities are too similar to within-blocks edge probabilities. 

The proof of the above theorem uses the property of the eigenvalue rigidity, i.e., the bulk eigenvalues of generalized Wigner matrix are not far from the corresponding bulk eigenvalues of the Gaussian Orthogonal Ensembles. 
In particular, we use this eigenvalue rigidty property together with the balanced block sizes (\ref{a1}) and the fact that within-block probability is greater than the between block probability to show that the second eigenvalue of at least one candidate Erd\"{o}s R\'{e}nyi block for underfitted model is considerably greater than $2$. This essentially means that our test can detect at least one non-Erd\"{o}s R\'{e}nyi block when $K < K^{(n)}_{\star}$.  And as $n$ increases, this signal becomes larger and therefore the asymptotic
power of the tests (\ref{ht})-(\ref{ht1}) under the model is underfitted goes to one.

\begin{corollary}\label{cons.K}\normalfont [Consistency of K]. 
The estimate obtained using the sequential procedure in Algorithm \ref{alg:1}, given in (\ref{hatk}), converges to the true number of communities (i.e., $K_{\star}^{(n)}$) provided the underlying SBM satisfies assumption \ref{a1} and additionally satisfies the following
conditions: i) $K^{(n)}_{\star} = O(n^{(1 - 3 \gamma)/(4 - 3 \gamma)})$ with the sparsity parameter satisfying $\rho_n = O(n^{-\gamma})$, ii) $\min_{1 \le K^{(n)}_{\star}}  G_{\rho_{n, K_{\star}}}(i,i) \ge (n/K^{(n)}_{\star})^{-2/3}$, iii) $\max_{1 \le i \le K^{(n)}_{\star}} G_{\rho_{n, K_{\star}}}(i,i) < 1/2$, and $\max_{1 \le i \le K^{(n)}_{\star}} G_{\rho_{n, K_{\star}}}(i,i) > 1/2 + \delta_0$  then $\hat{K}$ given in (\ref{hatk}) is consistent, i.e.,
\be
P(\hat{K} = K_{\star}^{(n)}) \to 1,
\ee
as $\alpha \to 0$ and $\delta_0$ is given in (\ref{delta0}).
\end{corollary}

\begin{proof}
The proof is given in the Supplement.
\end{proof}

As discussed before, Corollary \ref{cons.K} guarantees the consistency of the estimate obatined
from the sequential procedure in Algorithm \ref{alg:1}, i.e., $\hat{K}$ in (\ref{hatk}).
For showing the consistency of $\hat{K}$ in (\ref{hatk}), we had to show the following: i) The power of the sequential test in (\ref{ht})-(\ref{ht1}) when the model is underfitted (i.e., $K < K^{(n)}_{\star}$) goes to one, ii) The test statistics in (\ref{t.stat}) converges to the \emph{Tracy-Widom} distribution under the null (i.e., $K = K^{(n)}_{\star}$).
For showing the power converging to one, we use the eigenvalue rigidty property, the balanced block sizes (\ref{a1}), and the fact that within-block probability is greater than the between block probability to show that the second eigenvalue of at least one candidate Erd\"{o}s R\'{e}nyi block is considerably greater than $2$. Therefore, as $n$ increases the test statistics in (\ref{t.stat}) becomes large and the power goes to one.
For showing the convergence of the test statistics in (\ref{t.stat}) under the null (i.e., $\hat{K} = K_{\star}^{(n)}$), 
we assume that $\hat{g}$ is consistent and can recover the true community structure as $n \to \infty$. Recall from Subsection \ref{sec.consis} that we have several methods that can recover true community structure (i.e., $\hat{g}$ is consistent) when $\hat{K}= K^{(n)}_{\star}$  (i.e., under the null). In particular, the consistency of $\hat{g}$ under the null is a common assumption in 
the literature, which is used by many other methods such as \cite{lei16} and \cite{wang17}.

\subsection{Comparison with existing methods}
The main advantage of SMT over other existing methods is that it allows the number of
communities to increase with a rate of $O(n^{(1 - 3\gamma)/(4 - 3\gamma)})$ in sparse SBMs,
where $\rho_n = O(n^{-\gamma})$. \cite{lei16} proved the consistency of their approach when
the number of communities increased at the rate of $O(n^{1/6 - \tau})$ for arbitrarily small $\tau >0$
in dense SBM/DCSBM cases. However, their method does not extend to the sparse SBMs. 
BHMC and NB methods of \cite{can2015} allow the number of communities to depend on the
network size $n$ but the rate at which the number of communities increases with the network
size $n$ is not specified. Recall that \cite{can2015}'s BHMC and NB methods are only valid for a narrow 
range of average degree $d$, e.g., for Erd\'{o}s R\'{e}nyi graphs the average degree $d$
has to satisfy $d < n^{2/13}$. In summary, the other spectral methods such as \cite{lei16}, \cite{can2015} do not
provide a broad theoretical guarantee for sparse SBMs compared to SMT.
All the non-spectral methods assume the number of communities to be fixed.

Like other spectral methods, SMT is also computationally fast because it only requires computing the second eigenvalue 
of scaled adjacency matrices, which makes it convenient for estimating the number of communities
in large networks, such as scRNA-seq datasets.
Moreover, in numerical simulations, we observe that the performance of SMT is relatively better when the out-in ratio
(i.e. the ratio of between-block probability to within-block probability) is rather large.

\subsection{Model Selection}
The basic idea of SMT is to test whether a given block is an Erd\"{o}s R\'{e}nyi block.
Using this simple test, we proposed SMT for estimating the number of communities where the 
alternative is a finer composition of Erd\"{o}s R\'{e}nyi blocks. In general, our method can be adapted
to detect a variety of compositions of multiple Erd\"{o}s R\'{e}nyi blocks against other alternative 
models, such as overfitted SBMs, DCSBMs, or mixed membership stochastic block models, where the overfited models
refer to the models whose assumed number of communities accedes the true number of blocks.
This flexibility is advantageous for our method because Erd\"{o}s R\'{e}nyi blocks (or SBMs) can act as a null model for a more complicated network structure, in such scenarios our method can be used for comparing two competing models.
Table \ref{T.1} compiles the rejection rate under SMT under the null or when the alternative is an overfitted SBM or a DCSBM.
The rejection rate being near one in Table \ref{T.1} for the two alternatives shows the usefulness of SMT as a model selection tool.
However, for any model that is closer to the SBM model but not a SBM, the performance of SMT as a model selection tool will decrease.

\begin{table}[!hbt]
\caption{Power of SMT against overfitted SBMs and DCSBMs. We compiled the rejection rate for our method against the null (the true SBM), the overfitted SBM (a SBM with the number of communities as $K +1$ where $K$ is the true number of communities), and the DCSBM. All the models were generated with the network size and the average degree of the network fixed as $1000$ and $10$, respectively. The DCSBM was generated using $K$. The rejection rate for any scenario was computed by simulating  adjacency matrices $100$ times under the null, the overfitted SBM, and the DCSBM.
}\label{T.1}
\centering
\begin{tabular}{| l | l | l | l | }
\hline
True $K$ & Null & Overfitted SBM & DCSBM  \\
\hline
2 & 0.01 & 1 & 1 \\
3& 0 & 1 & 1\\
4& 0 & 1 & 1 \\
\hline
\end{tabular}
\end{table}

\subsection{Extension to Hierarchical Community Detection}
Hierarchical networks are popular because hierarchical networks go beyond simple clustering by explicitly
including organization at all scales in the network, simultaneously \citep{clauset2008}.
This essentially means that the number of communities in a network depends on the scale, i.e.,
at a higher resolution or finer scale the number of communities would be greater compared to a lower 
resolution or coarser scale. \cite{li2020} used this idea to estimate a large number of communities in a network.
In particular, they proposed a binary tree stochastic block model (a special case of the hierarchical model
discussed by \cite{clauset2008}) that recursively splits the communities into two at every scale unless a stopping
criterion is reached. Moreover, they argued that they can estimate a large number of communities in a network
provided we have a consistent stopping rule for selecting the resolution level. In this section, we show that SMT
can also be used as a consistent stopping rule in estimating a large number of communities in a hierarchical network. 
This is advantageous compared to using existing stopping criterion such as \cite{can2015}'s NB method, 
which are only theoretically valid for a narrow range of the average degree $d$ of the network. 

\subsection{Binary Tree Stochastic Block Model}
Following \cite{li2020}, let $S_{\omega} = \{0, 1 \}^{\omega}$ denote the set of all binary $\omega$ sequences. Then, every
community is identified by a unique binary string in the set $S_{\omega}$. Moreover, the total 
number of communities is given by the cardinality of $S_{\omega}$, i.e., $K = |S_{\omega}|$. For any node
$i \in \{1, \cdots, n \}$, let $c(i) \in S_{\omega}$ be the community label, and let $C_x =\{ i : c(i)= x\}$ consist of all nodes
in $\{1, \cdots, n \}$ that has the same community label $x$.

\begin{enumerate}\label{btsbm}
\item Let $G \in \mathbb{R}^{K \times K}$ be a matrix of probabilities defined by
\ba
G( \mathcal{I}(x), \mathcal{I}(x')) = q_{D(x, x')},
\ea
where $q_0, \cdots, q_{\omega}$ are arbitrary $\omega + 1$ parameters in $[0,1]$.

\item Let edges between all pairs of distinct nodes $i,j$ are independent Bernoulli with

\be\label{prob.BTSBM}
P(A(i,j) = 1) = G( \mathcal{I}(c(i)), \mathcal{I}(c(j)))
\ee

satisfying $Q = E(A)$.
\end{enumerate}

Following \cite{li2020}, let $Z \in \mathbb{R}^{n \times K}$ be the membership matrix with the $i^{th}$ row $Z_i =e_{\mathcal{I}(c(i))}$, where $e_{\mathcal{I}(c(i))}$ is the $\mathcal{I}(c(i))^{th}$ canonical basis vector in $\mathbb{R}^K$. Then,
the probability matrix can be given as
\be\label{p.rep}
Q = Z G Z^\top - q_0 I.
\ee

From (\ref{p.rep}), it is evident that $P$ is a matrix of rank $d$ that can be parametrized
using $(q_0, q_1, \cdots, q_d)$, where $q_l$ denotes the probability of forming an edge between any pair of nodes $i$ and $j$
depending on $l$ being the lowest level they share in the hierarchical network, where $0 \le l \le d$. For instance node $i$ and $i$ belong to the $0^{th}$ level (the lowest possible level), therefore the self-loop for node $i$ is generated with probability $q_0$.
\cite{li2020} showed that when the network is balanced then the eigenvalues of
$Q$ can be given in terms of block size and $(q_0, \cdots, q_i \cdots, q_d)$. The two most natural configurations arise
where a hierarchy is meaningful are \emph{assortative} communities satisfying $q_0 > \cdots > q_i > \cdots > q_d$
or \emph{dis-assortative} communities satisfying $q_0 < \cdots < q_i < \cdots < q_d$.
Also, $(q_0, \cdots, q_i, \cdots, q_d)$ could be reparametrized as below
\be\label{p_a}
 (q_0, \cdots, q_i, \cdots, q_d) = \rho_n(1, a_1, \cdots, a_d).
\ee

Several methods have been proposed for recovering the community structure in these settings. Essentially, these methods are
each a composition of a partitioning algorithm and a stopping rule that is applied recursively. A partitioning algorithm
partitions a network into sub-networks, whereas the stopping rule determines where the network should be
partitioned or not. This process is recursively applied until all subnetworks cannot be further partitioned (i.e., the stopping rule
rejects the further bipartition of any subnetworks). Following \cite{li2020}, we consider two bipartition 
algorithms, namely: Simple eigenvector sign check algorithm (SES) in Algorithm \ref{alg:2} and 
Regularized spectral clustering algorithm (RSC) in Algorithm \ref{alg:3}.

\begin{algorithm}
\caption{Simple Eigenvector Sign Check (SES)}
\label{alg:2}
\begin{algorithmic}[1]
\State Given any adjacency matrix A, compute the eigenvalue $\hat{v}_2$ correspondng to the second largest eigenvalue
\State Let $\hat{c}(i)=0$ if $\hat{v}_2 \ge 0$ and $\hat{c}(i)=1$ otherwise.
\State Return label $\hat{c}$.
\end{algorithmic}
\end{algorithm}

\begin{algorithm}
\caption{Regularized Spectral Clustering (RSC)}
\label{alg:3}
\begin{algorithmic}[1]
\State Input any adjacency matrix $A$ and a regularization parameter $\tau$ with the default value of $0.1$.
\State Compute the regularized adjacency matrix as
\ba
A_{\tau} = A + \tau \frac{\bar{d}}{n} \boldsymbol{1}\boldsymbol{1}^\top,
\ea
where $\bar{d}$ is the average degree of the network.
\State Let $D_{\tau}= diag(d_{\tau 1}, \cdots, d_{\tau n})$, where $d_{\tau i} = \sum_{j} A_{\tau, ij}$. Then, the regularized
Laplacian is given as below
\ba
L_{\tau} = D_{\tau}^{-1/2} A_{\tau} D_{\tau}^{-1/2}.
\ea
\State Compute the leading two eigenvalues of $L_{\tau}$ and arrange them in a $n \times 2$ matrix $U$, and then
$K-means$ algorithm with $K=2$ gives the required two way partition.
\end{algorithmic}
\end{algorithm}

\cite{li2020} showed the above approach  can result in the recovery of underlying community structure as long
as the stopping rule is consistent. It is intuitive to see that the stopping rule is crucial for determining the
correct depth (or resolution) of the hierarchical network and therefore inconsistent stopping rule may not
yield in the recovery of true community structure. \cite{li2020} gave the following definition of
the consistency of the stopping rule. 

\begin{definition}[Stopping Rule]\label{stop.rule}
A stopping rule for a network of size $n$ generated from an SBM with $K$ communities is consistent with the
rate $\phi$ if $P(\psi(A) = 1) \ge 1 - n^{-\phi}$ when $K \ge 1$ and $P(\psi(A)=0) \ge 1 - n^{-\phi}$ when $K=1$.
\end{definition}

The stopping rule in the case of BTSBM determines whether a given subnetwork
is a SBM with block size $K=1$ (i.e., an Erd\"{o}s R\'{e}nyi block). 
\cite{li2020} recommended using NB method as the stopping rule.
From the discussion in Section \ref{s.main}, it is evident that we can use SMT
with a nominal significance level $\alpha$ for the stopping role (i.e., SMT for
testing whether a network or a subnetwork is an Erd\"{o}s R\'{e}nyi block or not).
The consistency of using SMT as a stopping rule follows directly from Corollary \ref{cons.K}. 
Theoretically, SMT has a slight advantage over the NB method for being used as a stopping rule
because SMT is consistent for a broader range of the sparsity parameter.
Moreover, for \emph{associative} and \emph{dis-associative} hierarchy under the balanced assumption \ref{a1},
 it follows from \cite{li2020} (see Theorems $2$ and $3$) that a recursive bipartition method 
paired with any consistent stopping rule, such as SES with SMT (SES.SMT)
can consistently recover true community structure of the BTSBM.

\section{Numerical Experiments}\label{s.num} We perform three separate
numerical experiments to evaluate the performance of our method. 
The first and second numerical experiment compares the performance
of SMT on SBMs with a small number of communities and a large number of communities, respectively.
The third numerical experiment evaluates the performance of community extraction with SMT as
stopping rule (SES.SMT) for binary tree stochastic block models.

For the first two experiments we generated adjacency matrices A according to the SBM model:
\ba\label{num.mod}
& A_{ij}\mid (g_i,g_j) \sim Ber(1, \rho_n d +\rho_n s* I\{ g_i =g_j \}), i=1,\cdots, n,
g_i=1,\cdots, K,
\ea
where $g=(g_1,\cdots, g_n)$ is the community membership vector, $Ber(\cdot)$ denotes Bernoulli distribution, $I(\cdot)$
denote the indicator function, $s\rho_n$ denotes the between-block
edge probability, $d\rho_n$ is the difference of
within-block and between-block edge probabilities,
the out-in ratio is the ratio of between-block probability to within-block probability (i.e. $\frac{d}{d + s}$), and $K$ is the
true number of communities.

\begin{figure}
\centering
\includegraphics[width=15.5cm]{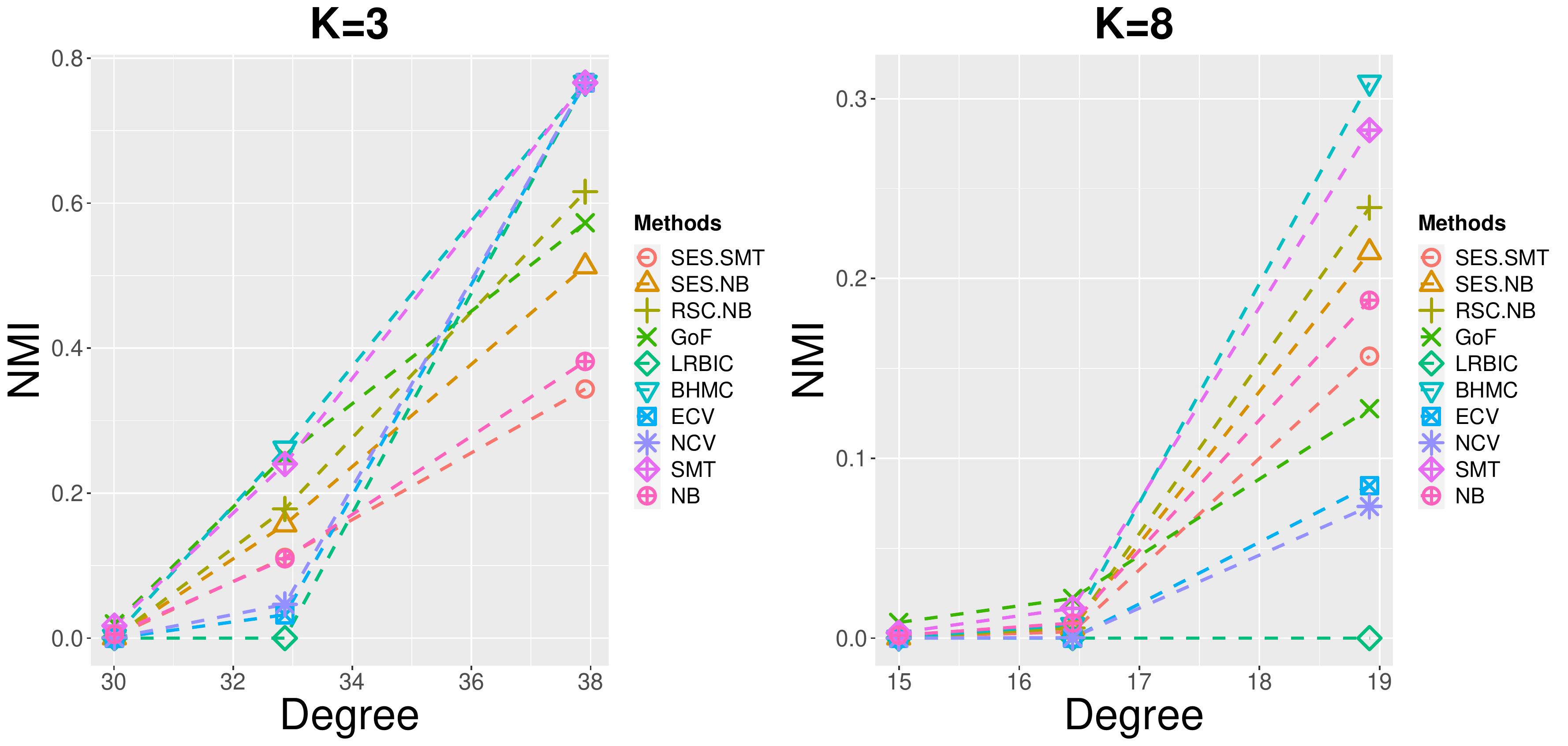}
\caption{NMI comparison of methods for estimating the number of communities in SBMs at a high out-in ratio. We compiled  normalized mutual information across different methods for estimating number of communities in SBMs of network size of $1000$ nodes with the
true number of communities $K$ varying over $(3,8)$ while keeping out-in ratio as $0.40$.}\label{F:comp_exhaust}
\end{figure}

The first numerical experiment is conducted on a SBM of size $1000$ with a small number
of communities. The methods that do well in this experiment are considered for the more computationally expensive
second experiment. Therefore for the first experiment, we select parameters such that they are discriminating.
In particular, we chose the out-in ratio as $0.4$ while we varied the true number of communities 
$K$ over $(3, 8)$. Increasing the out-in ratio negatively affects all methods. 
For studying the performance of methods with increasing $K$, we keep the out-in ratio fixed at $0.4$.
Moreover, the sparsity parameter for this experiment was $\rho_n=N^{-1/6}$ and the
within block probability is varied over $(0.05, 0.1, 0.15, 0.2)*\rho_n$. We evaluated every method
using normalized mutual information (NMI) that compares the estimated cluster labels with the true cluster
label with $NMI$ varying from $0$ to $1$. The higher values correspond to the better quality of clustering and
the better accuracy of methods that estimate the number of communities.
For every scenario, we generated the adjacency matrix $100$ times and then compiled the 
NMI of various methods in Figure \ref{F:comp_exhaust}. 
For SMT, NMI is fairly robust to the choice of significance level and so we kept the
signifcance level as $0.05$.

Figure \ref{F:comp_exhaust} suggests that BHMC and SMT are two stand-out methods in the first
experiment.  For small $K$, we observe that the performance of non-spectral methods are more 
sensitive to the changes in the degree compared to the non-spectral methods or even hierarchical methods.
This is because the spectral  methods utilize few eigenvalues to estimate the number of communities, which
tend to be more robust with changes in degrees.
For large $K$, we see that the performance of non-spectral methods is underwhelming even
for higher degrees. This is expected because the non-spectral methods assume the
number of communities to be fixed. Figure \ref{F:comp_exhaust} also suggests that the hierarchical
methods have somewhat middling performance.

For the second experiment, we compared the performance of SMT against BHMC and SES.NB, SES.SMT and RSC.NB.
We chose hierarchical methods because they had somewhat better performance than the non-spectral
methods for large $K$ and the ability of hierarchical methods to estimate a large number of communities, see \cite{li2020}.
Like in the first experiment, we generated adjacency matrices $A$ from SBM using (\ref{num.mod}).
In particular, we simulated multiple networks of size $n=5000$ with the fixed out-in ratio as 
$1/3$ and $0.4$ while we varied the true number of communities over $(10, 20, 30, 60)$ and the degree over
$(50, 100, 150, 200)$. Like in the first experiment for every scenario, we generated $100$ copies of 
adjacency matrices from the SBM and then compiled the normalized mutual information (NMI) of various methods.
As discussed before, we kept the significance level as $0.05$.
Table \ref{T.2} compiles the normalized mutual
information (NMI) for all the four methods when the out-in ratio was $1/3$. It is immediate that as the true number
of communities increases the performance of every method decreases. However, SMT has comparatively better 
performance in all scenarios. Table 2 in the Supplement gives a similar comparison of results when the out-in ratio
was $0.4$. In this scenario too, SMT has a comparatively better performance than the rest of the methods.
The slightly improved performance of SMT over BHMC in Table \ref{T.2} compared to Figure \ref{F:comp_exhaust}
can be attributed to the improved power of SMT as the network size increases. Specifically, in underfitted models
as the network size increases one of the candidate eigenvalues corresponding to the candidate Erd\"{o}s R\'{e}nyi blocks
becomes large enough under the assumption of balanced block sizes (\ref{a1}).

\begin{table}[!hbt]
\caption{NMI comparison of methods for estimating the number of communities in SBMs.
We compiled the normalized mutual information for estimating the number of communities in networks of size $5000$ for multiple scenarios with the true number of communities ($K$) varying over $(10,20,30,60)$ and the average degree varying over
$(50, 100, 150, 200)$ and the out-in ratio was $1/3$.}\label{T.2}
\centering
\begin{tabular}{| l | l | l | l | l | l | }
\hline
True $K$ & d  & SES.NB & RSC.NB & SMT & BHMC \\
\hline
10 & 50  & 0.242 & 0.641 & 0.946 & 0.848 \\
20 & 50  & 0.082 & 0.11 & 0.316 & 0.193 \\
30 & 50  &  0.006 & 0.006 & 0.028  & 0.007  \\
60 & 50  & 0 & 0 & 0.004 & 0 \\
\hline
10 & 100 & 0.384 & 0.921 & 0.996 & 0.935 \\
20 & 100 & 0.169 & 0.524 & 0.954 & 0.440 \\
30 & 100 & 0.089 & 0.145 & 0.477 & 0.185 \\
60 & 100 & 0 & 0 & 0.006 & 0  \\
\hline
10 & 150 & 0.472 & 0.972 & 0.996 & 0.940 \\
20 & 150 & 0.229 & 0.791& 0.980 & 0.51  \\
30 & 150 & 0.135 & 0.352 & 0.922 & 0.290  \\
60 & 150 & 0 & 0 & 0.042 & 0 \\
\hline
10 & 200 & 0.522 & 0.984 & 0.995 & 0.938 \\
20 & 200 & 0.273 & 0.883 & 0.979  & 0.528  \\
30 & 200 & 0.169 & 0.606 & 0.976 & 0.343  \\
60 & 200 & 0.018 & 0.024 & 0.153 & 0.011  \\
\hline
\end{tabular}
\end{table}

For the third scenario, we generated adjacency matrix $A$ according to the BTSBM model in (\ref{btsbm}).
For the third experiment, we compared the extension of SMT to hierarchical networks against competing
methods for hierarchical networks. For this experiment, we generated adjacency matrices from BTSBM.
In this experiement, we varied the depth of the network over $(2, 4, 6, 8)$ while varying the average
degree of the network. Moreover, we fixed hierarchical probabilities $(p_0, \cdots, p_d) =(a^0, \cdots, a^d)$.
As discussed before, we used normalized mutual information (NMI) for comparing
the performance of all the methods. Table \ref{T.3} compiles the performance of our method.
It is evident from Table \ref{T.3} that SES.SMT has sub-par performance for small value of true number
of communities $K$, but it catches up as the true number of communities $K$ increases to $256$.
This is largely because the performance of other methods tapers off as the true number
of communities increases.

\begin{table}[!hbt]
\caption{NMI comparison of methods for estimating the number of communities in BTSBM.
We compiled the normalized mutual information for estimating the number of communities in networks of size $6400$ for multiple scenarios with the true number of communities ($K$) varying over $(4, 16, 64, 256)$ and $a$ varying over
$(0.2, 0.4)$ .}\label{T.3}
\centering
\begin{tabular}{| l | l | l | l | l |   }
\hline
True $K$ & a & SES.SMT & SES.NB & RSC.NB \\
\hline
4 & 0.2  & 0.5   & 1.000 & 1.000 \\
16 & 0.2 & 0.750  & 1 & 1 \\
64 & 0.2 & 0.833  & 1.00 & 1.000 \\
256 & 0.2 & 0.866 & 0.866 & 0.874  \\
\hline
4 & 0.4 & 0.503   & 1.000 & 1.000 \\
16 & 0.4 & 0.751  & 0.999 & 0.999 \\
64 &  0.4& 0.833  & 1.00 & 1.000 \\
256 & 0.4 & 0.875 & 0.875 & 0.875  \\
\hline
\end{tabular}
\end{table}

\section{Real Data Analysis}\label{s.data}
In this subsection, we perform two types of real data analysis.
First, we benchmark the performance of SMT against other competing
methods on six benchmark scRNA-seq datasets for which the reference
clusters are known. Second, we use SMT and HCD-SMT to estimate the
clusters in a sparse scRNA-seq datasets.

\subsection{Benchmark Data Analysis}

For benchmark analysis, we run our comparison analysis on six scRNA-seq datasets that were considered
\emph{gold-standard} in \cite{kiselev2017}. The six reference datasets can be downloaded
from Gene Expression Omnibus (GEO) \cite{geo} with their ascension given in Table \ref{T.5}.
For comparison analysis, we run SMT, SC3 \citep{SC3}, SIMLR \citep{SIMLR}, and SEURAT \citep{satija2015}
on the six reference datasets. Before running the comparison analysis,
we first run data preprocessing steps on the six reference datasets.
Subsequently, we processed the scRNA-seq data in line with the current 
best practices in the existing literature, see \cite{best_practice_scRNA}, \cite{SC3}.
The data preprocessing steps were common for all the four methods.
In the data preprocessing step, we discarded genes that have low variability. In particular, we kept
genes whose variability is at least greater than the $50^{th}$ quantile. Then, we normalized the
remaining single-cell data using the $log_2(1 + x/10000)$ transformation. Subsequently, we ran
the rest of the analyses for SC3, SEURAT, and SIMLR using the default parameter settings.
Since the filtered benchmark data was fairly dense, therefore for running SMT we generated 
an adjacency matrix $A$ using the correlation matrix of transformed single-cell data with the
$50^{th}$ quantile as the cut-off for forming an edge.

\begin{table}[!hbt]
\caption{Summary of six reference datasets. The following is the summary level information on six reference datasets along with
their GEO ascension numbers and the original papers.}\label{T.5}
\centering
\begin{tabular}{| l | l | l | l | l | l |  }
\hline
Datasets & Number & Number & Cell Resource & GEO ascension  & Reference Paper \\
& of Cells & of Genes &  & number&\\
\hline
Biase & 49 & 25,737 & 2-cell and 4-cell & GSE57249 & \cite{biase2014}\\
 & &  &  mouse embryos &  & \\
Yan & 124 & 22,687 & Human preimplantation   & GSE36552 & \cite{yan2013} \\
& & & embryos and & & \\
& & & embryonic stem cells & & \\
Goolam & 124 & 41,480 & 4-cell mouse embryos  & E-MTAB-3321 & \cite{goolam2016} \\
 & &  & embryos  & &  \\
Deng & 268 & 22,457 & Mammalian cells & GSE47519 &  \cite{deng2014} \\
Pollen & 301  & 23,730 & Human cerebral cortex & SRP041736 & \cite{pollen2014} \\
Kolodziejczyk & 704 & 38,653 & Mouse embryonic & E-MTAB-2600 & \cite{kolodziejczyk2016} \\
 & &  & stem cells &  &  \\
\hline
\end{tabular}
\end{table}

\begin{figure}
\caption{Visual representation of the adjusted rand index computed on reference single-cell datasets.
First, we leave out genes whose variability is less than the $50^{th}$ quantile. Second, we normalize
the single-cell data by dividing it by $1000$. Third, we transform the single cell data using $log(1+ x)$
function. Fourth,  we used SMT, Seurat, SC3, and
SIMLR to obtain the estimated cell clusters of the six reference datasets, namely: Biase, Deng, Goolam,
Kolodziejcky, Pollen, and Yan. Finally, we used original cell clusters (obtained by the respective
authors) and the estimated cell clusters to obatin the adjusted rand index for all four methods.
}\label{F:ARI}
\includegraphics[width=5in]{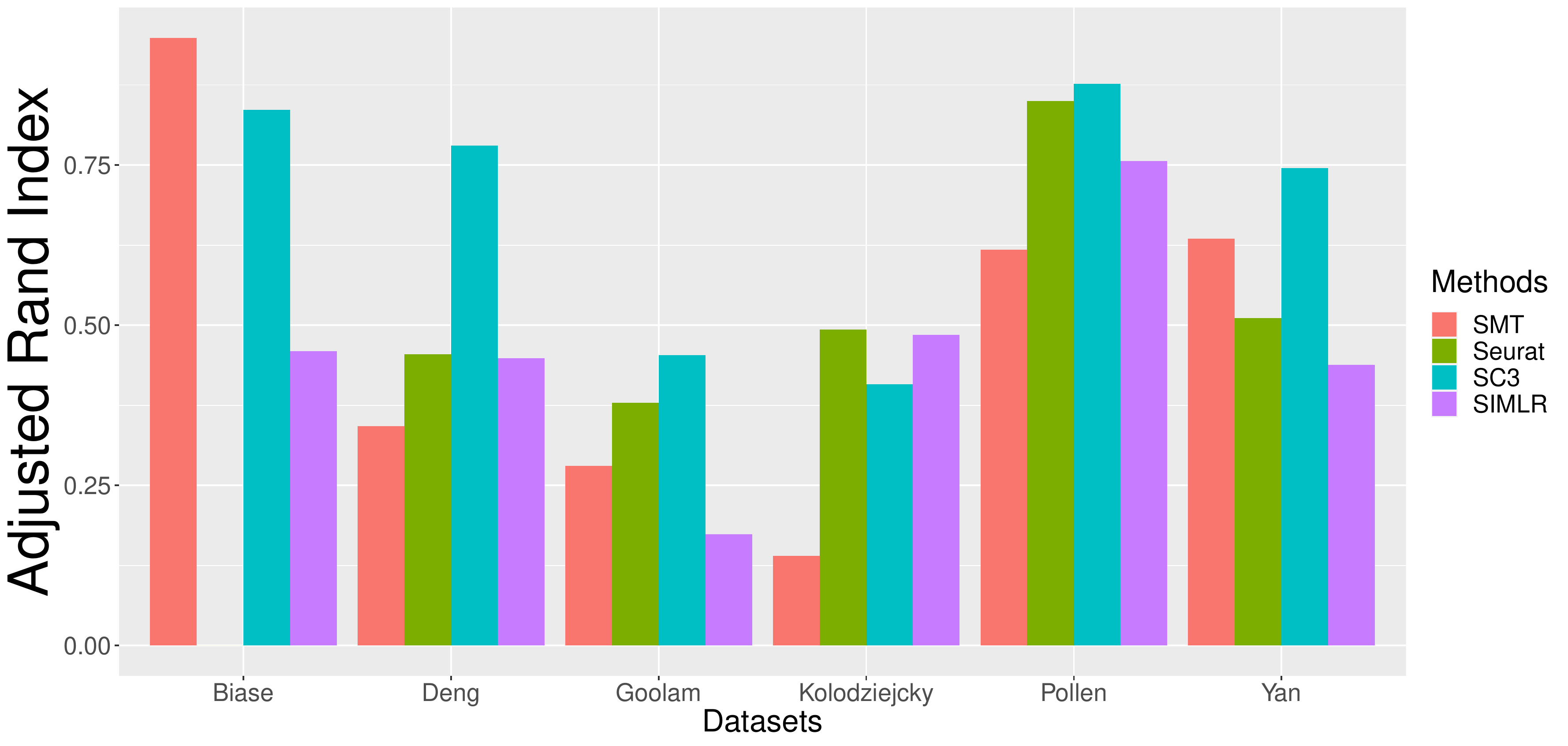}
\end{figure}

Post preprocessing, we run SMT, SC3, SEURAT, and SIMLR on the six reference datasets to get estimated
cell cluster labels for all the four methods. For SMT, we chose the significance level $\alpha =0.05$.
Using the reference cell cluster labels for the six 
reference datasets and the estimated cell cluster labels, we compute adjusted rand index (ARI) for all four
methods across six reference datasets (see Figure \ref{F:ARI}). Table \ref{T.4} gives the estimated cell cluster
across the six reference datasets along with the reference number of cell clusters.

Overall, Figure \ref{F:ARI} suggests that SC3 performs well across the six reference datasets, while 
SMT has middling performance. Table \ref{T.4} also suggests the same. The performance of SMT is
sensitive to SBM assumptions and the choice of $50^{th}$ quantile in forming an edge in adjacency
matrices. Selecting the $50^{th}$ quantile is akin to assuming that the network is dense.
The rationale for dense network stems from the fact that our pre-processing has selected only
the top half of most variable genes and therefore it is safer to assume that filtered scRNA-seq data have
less sparsity. However, it is worth noting that increasing the cut-off would increase the sparsity
 and negatively impact the performance of SMT. Table in the Supplement compares
the performance of SMT when the cut-off used is the $95^{th}$ quantile instead of the $50^{th}$ quantile.

\begin{table}
\caption{Comparison table for estimated number of communities for reference single cell datasets.}\label{T.4}
\begin{tabular}{| l | l | l | l | l | }
\hline
Datasets & Reference & SMT  & SC3 & SIMLR \\
\hline
Biase & 3 & 3  & 3 & 7 \\
Yan & 7 & 3  & 6 & 12 \\
Goolam & 5 & 10  & 6 & 19  \\
Deng & 10 & 10  & 9 & 16 \\
Pollen &11 & 11  & 11 & 15\\
Kolodziejcky & 3 & 27  & 10 & 6\\
\hline
\end{tabular}
\end{table}

\subsection{Sparse Single-cell Data Analysis}
For the rest of the data analysis, we use scRNA-seq data generated from the retina
cells of two healthy adult donors. 
The scRNA-seq data was generated from the retina cells of two healthy
adult donors using the 10X Genomics Chromium$^{TM}$ system. Detailed
preprocessing and donor characteristics of our scRNA-seq data can be found
in \cite{lyu2019} (see Supplementary Note 1).
In total, 33694 genes were sequenced over 92385 cells.
The sequencing data were initially analyzed with R package Seurat (\cite{satija2015}) and every 
cell was identified as a particular cell type.
Table 3 in the supplementary material gives the relative size of these cell types. 

For sparse single-cell data analysis, we consider two types of data analysis: i) Composite data analysis,
ii) Subgroup analysis of bipolar cells. In composite data analysis,
we combine different single cells with classification known from Seurat and then we use SMT
to retrieve the estimated number of cell clusters.
The composite analysis aimed to see if we can recover the true number of cell clusters when
the number of true cell clusters was well-known in advance.
In the subgroup analysis, we use the hierarchical version of SMT (SES.SMT) for the clustering of bipolar cells.
The rationale for clustering bipolar cells using SES.SMT is that the bipolar cells have multiple subgroups
with the possibility of having a hierarchical structure.

\begin{figure}[!hbt]
\centering
\includegraphics[width=4in]{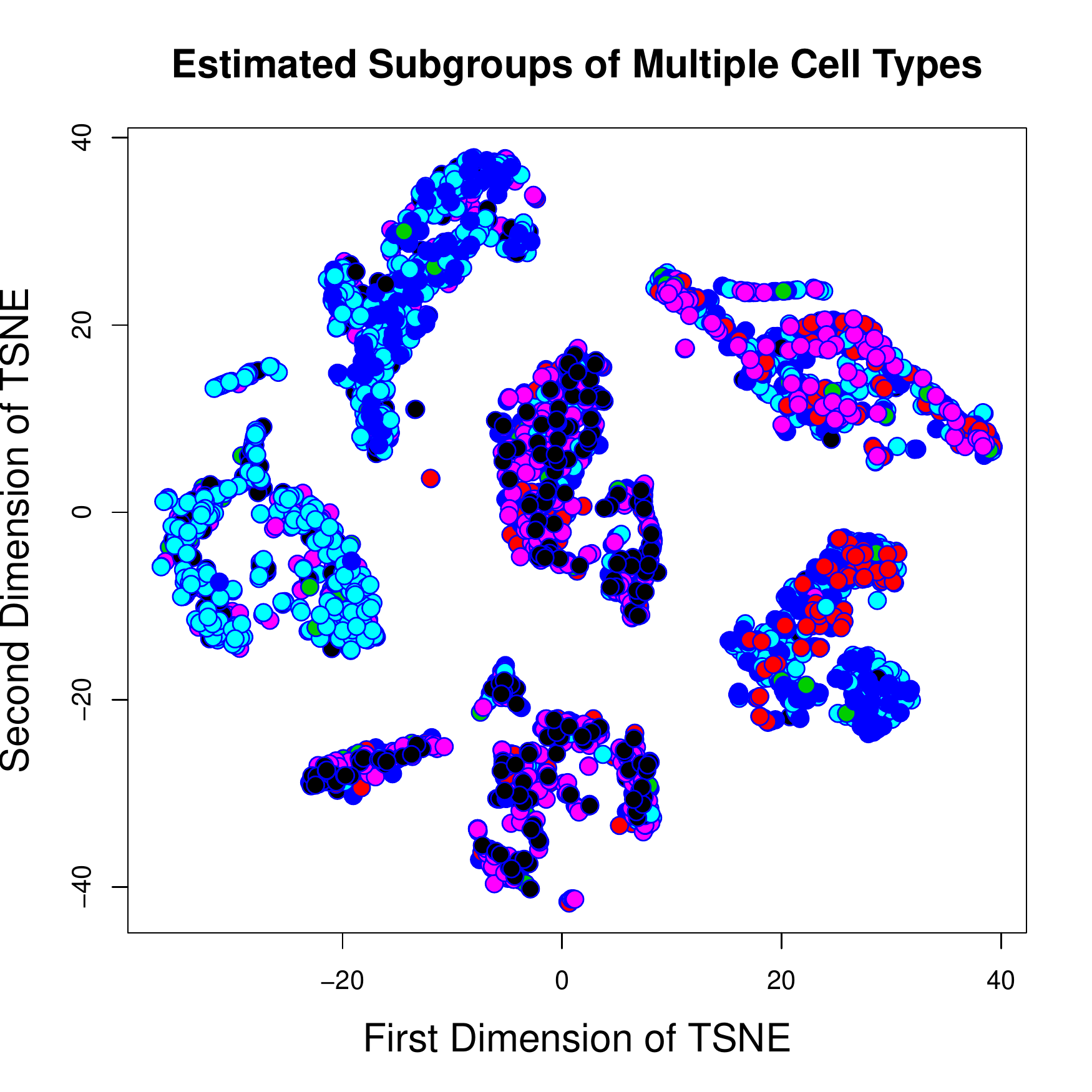}
\caption{Visual representation of the estimated clusters of the composite scRNA-seq data of human retina cells in the t-SNE plot.
The composite scRNA-seq data was obtained from the scRNA-seq data of human retina cells by selecting
Seurat classified cell types namely: Ganglion, Endothelium, Cone, Bipolar, Horizontal, 
and Rod cells in an equal manner of roughly $500$ cells per cell type. The composite network data was generated
by assigning the edge between a pair of cells whenever the correlation between the two cells was greater
than the $(1 - \rho_n)^{th}$ quantile of the elements of the sample correlation matrix, where $\rho_n =0.05$. Subsequently, using SMT, we obtained the estimate of the number of communities in the composite network data as six. Also, we used the sample correlation matrix to generate the t-SNE plot. The estimated clusters are shown using different colors
in the t-SNE plot.
}\label{F:All}
\end{figure}

For composite data analysis, we analyzed subsets of Ganglion, Endothelium, Cone, Bipolar, Horizontal, and  Rod cell types.
The composite network was obtained by combining equal samples of size $500$ from each of the above cell types. 
For the above network, we used correlations between cells to compute similarity between cells. 
Subsequently, we used the correlation matrix with $95^{th}$ quantile (of entries of the correlation matrix) as the cutoff to generate
an adjacency matrix. Subsequently, we used SMT to estimate the number
of communities. The rationale for running SMT was the composite data was artifically constructed as composition
of six different cell types. SMT and BHMC estimated the number of communities as six whereas LRBIC, ECV, NCV estimated the number of communities as $25$, $29$, and $29$ respectively. It is evident that both SMT and BHMC recovered the true number of cell clusters
in the composite network. Figure \ref{F:All} gives the t-SNE plot for the estimated composite network.

\begin{figure}
\includegraphics[width=4 in]{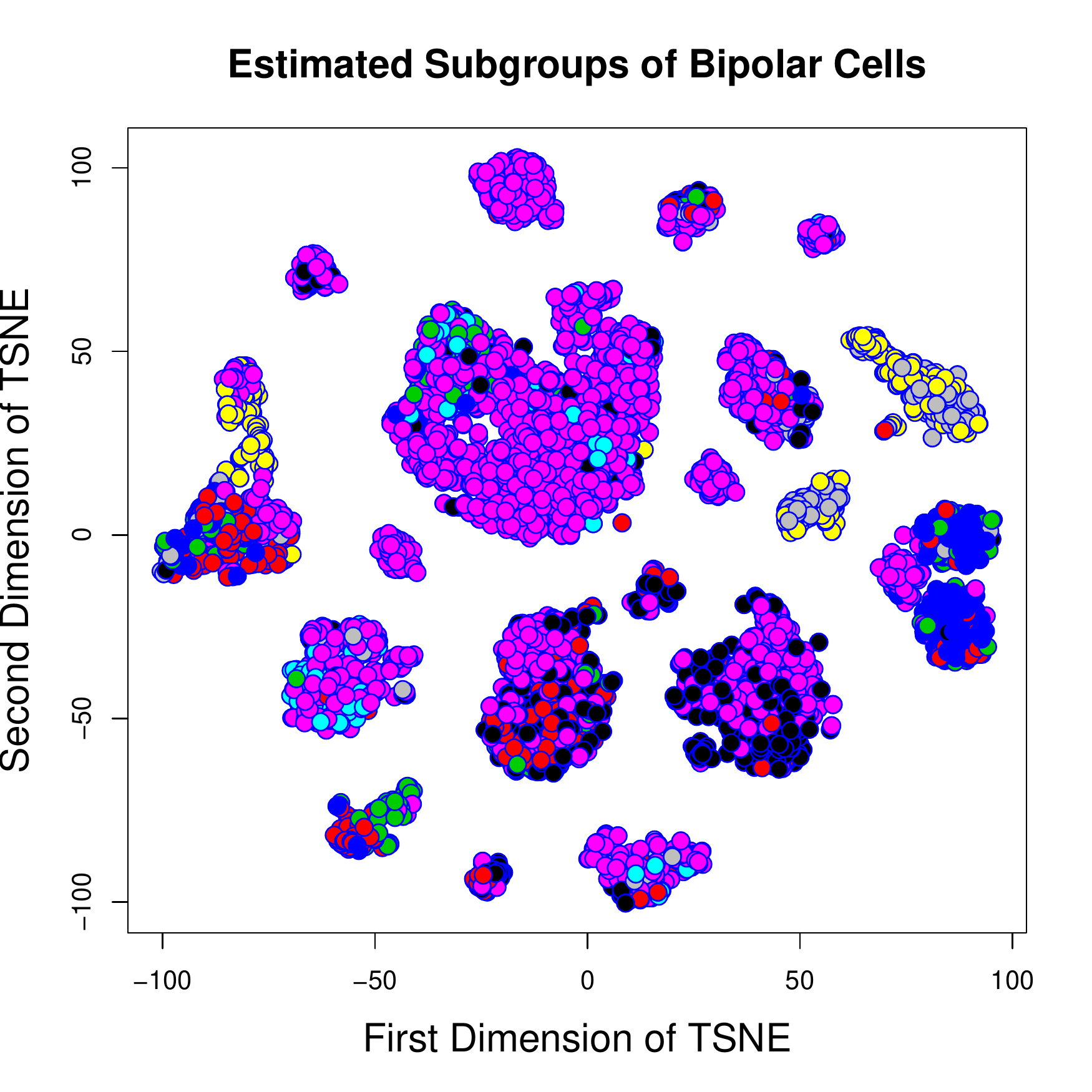}
\caption{Visual representation of estimated subgroups of bipolar cell types of human retina single-cells in the t-SNE plot.
The bipolar single cells from the original scRNA-seq of retina cells were selected using the Seurat classified cell types.
The  bipolar cells were then filtered and normalized using the general data manipulating strategies for scRNA-seq datasets. 
Subsequently, we computed the sample correlation matrix and the network data was generated by assigning an undirected edge between a pair of cells when the correlation between the two cells was greater than the $95^{th}$ quantile of the entries of the sample correlation matrix. Then, we used the hierarchical extension of SMT (i.e., SES.SMT) to estimate the number of communities. The estimated subgroups are denoted by different colors in the t-SNE plot.}\label{F:Retina}
\end{figure}

Our scRNA-seq data of bipolar cells was given in a matrix form with genes denoting the rows and cells denoting the columns.
In total, we had $33,694$ genes and $30,125$ cells. Subsequently, we processed the scRNA-seq data in line with the current 
best practices in the existing literature, see \cite{best_practice_scRNA}, \cite{SC3}. In particular, we filtered out genes whose variability was less than the $50^{th}$ quantile. And we filtered out cells whose total cell counts (across all genes) were less than $500$ and greater than $2500$. Subsequently, we normalized the data using $log_2(1 + x/10000)$. Then, we computed the correlation matrix between the cells. Using the correlation matrix and the $95^{th}$ quantile of entries of the sample correlation matrix as the cutoff to generate the adjacency matrix. We used the cut-off $95^{th}$ quantile of the entries of the sample correlation matrix to keep the sparsity in the range of theoretical value of the sparsity parameter for which SMT holds. Then, we used the hiearchical version of SMT (i.e., SES.SMT) to estimate the number of communities. The rationale for selecting SES.SMT was that the bipolar
cell types are believed have to number of subgroups with possiblly hiearchical structure.
Figure \ref{F:Retina} plots the estimated subgroups of single cells along the two dimensions of t-SNE. The estimated subgroups in the figure appear as clustered with a hierarchy.
The comparison of highly expressed genes of the subgroups of bipolar cells of a healthy patient against
the highly expressed genes of the subgroups of bipolar cells of a diseased patient can potentially help
uncover driver genes \citep{diffgenes2015}.

\section{Discussion}\label{s.disc}
In summary, SMT is useful for estimating the number of communities in sparse and large
SBMs. Moreover, SMT can be adapted as a stopping rule for BTSBMs. The main advantage of
SMT over other competing approaches is it has broad theoretical guarantees for sparse SBMs 
while allowing the number of communities to increase with the network size. This has wider
implications for application areas such as clustering of large scRNA-seq datasets where assuming a fixed number of
communities could be limiting. Moreover, SMT can estimate the number of communities 
increasing at the order of $O(n^{1/4})$, which is much higher than the $O(n^{1/6})$ that the GoF
method of \cite{lei16} can estimate. 

A drawback of SMT is that it does not automatically extend to the DCSBMs. The main argument
behind this assertion is that SMT uses the fact that all but the top eigenvalues of appropriately
scaled adjacency matrices under the null converges to the semi-circular 
distribution which is not necessarily true for the corresponding scaled adjacency matrices for the blocks/communities
of DCSBMs. There is a possibility to have a SMT-like procedure for estimating the number of blocks in DCSBMs provided that
we could characterize non-homogenous Erd\"{o}s R\'{e}nyi block in terms of its spectral properties.

\vskip 14pt
\noindent {\large\bf Supplementary Materials}
The Supplement includes proofs and additional information related to real datasets used for the analysis.

\par

\bibliographystyle{chicago}
\bibliography{Aoas_SMT_v1}

\begin{thebibliography}{}

\bibitem[\protect\citeauthoryear{Airoldi, Blei, Fienberg, and Xing}{Airoldi
  et~al.}{2008}]{airoldi2008}
Airoldi, E.~M., D.~M. Blei, S.~E. Fienberg, and E.~P. Xing (2008).
\newblock Mixed membership stochastic blockmodels.
\newblock {\em Journal of Machine Learning Research\/}~{\em 9}, 1981--2004.

\bibitem[\protect\citeauthoryear{Amini, Chen, Bickel, and Levina}{Amini
  et~al.}{2013}]{amini2013}
Amini, A.~A., A.~Chen, P.~J. Bickel, and E.~Levina (2013).
\newblock Pseudo-likelihood methods for community detection in large sparse
  networks.
\newblock {\em Annals of Statistics\/}~{\em 41\/}(4).

\bibitem[\protect\citeauthoryear{Amini and Levina}{Amini and
  Levina}{2018}]{amini2018}
Amini, A.~A. and E.~Levina (2018).
\newblock On semidefinite relaxations for the block model.
\newblock {\em Annals of Statistics\/}~{\em 46\/}(1), 149--179.

\bibitem[\protect\citeauthoryear{Balakrishnan, Xu, Krishnamurthy, and
  Singh}{Balakrishnan et~al.}{2011}]{balakrishnan2011}
Balakrishnan, S., M.~Xu, A.~Krishnamurthy, and A.~Singh (2011).
\newblock Noise thresholds for spectral clustering.
\newblock {\em Advances in Neural Information Processing Systems\/}, 954--962.

\bibitem[\protect\citeauthoryear{Biase, Cao, and Zhong}{Biase
  et~al.}{2014}]{biase2014}
Biase, F.~H., X.~Cao, and S.~Zhong (2014).
\newblock Cell fate inclination within 2-cell and 4-cell mouse embryos revealed
  by single-cell rna sequencing.
\newblock {\em Genome Res.\/}~{\em 24}, 1787--1796.

\bibitem[\protect\citeauthoryear{Bickel and Chen}{Bickel and
  Chen}{2009}]{bickel09}
Bickel, P.~J. and A.~Chen (2009).
\newblock A nonparametric view of network models and newman-girvan and other
  modularities.
\newblock {\em Proceedings of the National Academy of Sciences of the United
  States of America\/}~{\em 106\/}(50), 21068--21073.

\bibitem[\protect\citeauthoryear{Blondel, Guillaume, Lambiotte, and
  Lefebvre}{Blondel et~al.}{2008}]{blondel2008}
Blondel, V.~D., J.-L. Guillaume, R.~Lambiotte, and E.~Lefebvre (2008).
\newblock Fast unfolding of communities in large networks.
\newblock {\em Journal of Statistical Mechanics Theory and Experiment\/}.

\bibitem[\protect\citeauthoryear{Chaudhuri, Chung, and Tsiatas}{Chaudhuri
  et~al.}{2012}]{chaudhuri2012}
Chaudhuri, K., F.~Chung, and A.~Tsiatas (2012).
\newblock Spectral clustering of graphs with general degrees in the extended
  partition model.
\newblock {\em JMLR : Workshop and Conference Proceedings vol\/}, 35.1--35.23.

\bibitem[\protect\citeauthoryear{Chen and Lei}{Chen and Lei}{2018}]{chen2018}
Chen, K. and J.~Lei (2018).
\newblock Network cross-validation for determining the number of communities in
  network data.
\newblock {\em Journal of the American Statistical Association\/}~{\em
  113\/}(521).

\bibitem[\protect\citeauthoryear{Choi, Wolfe, and Airoldi}{Choi
  et~al.}{2012}]{choi2012}
Choi, D.~S., P.~J. Wolfe, and E.~M. Airoldi (2012).
\newblock Stochastic blockmodels with a growing number of classes.
\newblock {\em Biometrika\/}~{\em 99\/}(2), 273--284.

\bibitem[\protect\citeauthoryear{Clauset, Moore, and Newman}{Clauset
  et~al.}{2008}]{clauset2008}
Clauset, A., C.~Moore, and M.~E.~J. Newman (2008).
\newblock Hierarchical structure and the prediction of missing links in
  networks.
\newblock {\em Nature\/}~{\em 453}, 98--101.

\bibitem[\protect\citeauthoryear{Deng, Ramsk\"{o}ld, Reinius, and
  Sandberg}{Deng et~al.}{2014}]{deng2014}
Deng, Q., D.~Ramsk\"{o}ld, B.~Reinius, and R.~Sandberg (2014).
\newblock Single-cell rna-seq reveals dynamic, random monoallelic gene
  expression in mammalian cells.
\newblock {\em Science\/}~{\em 343}, 193--196.

\bibitem[\protect\citeauthoryear{Ding, Zhang, and Luo}{Ding
  et~al.}{2016}]{ding2016}
Ding, Z., X.~Zhang, and B.~Luo (2016).
\newblock Overlapping community detection based on network decomposition.
\newblock {\em Scientific Reports\/}~{\em 6\/}(24115).

\bibitem[\protect\citeauthoryear{Eberwine, Jai-Yoon, Bartfai, and Kim}{Eberwine
  et~al.}{2014}]{eberwine2014}
Eberwine, J., S.~Jai-Yoon, T.~Bartfai, and J.~Kim (2014).
\newblock The promise of single cell sequencing.
\newblock {\em Nature Methods\/}~{\em 11\/}(1), 25--27.

\bibitem[\protect\citeauthoryear{Edgar, Domrachev, and A.E.}{Edgar
  et~al.}{2002}]{geo}
Edgar, R., M.~Domrachev, and L.~A.E. (2002).
\newblock Gene expression omnibus: Ncbi gene expression and hybridization array
  data repository.
\newblock {\em Nucleic Acids Res.\/}~{\em 30\/}(1), 207--210.

\bibitem[\protect\citeauthoryear{Gao, Ma, Zhang, and Zhou}{Gao
  et~al.}{2017}]{gao2017}
Gao, C., Z.~Ma, A.~Y. Zhang, and H.~H. Zhou (2017).
\newblock Achieving optimal misclassification proportion in stochastic block
  models.
\newblock {\em The Journal of Machine Learning Research\/}~{\em 18\/}(1),
  1980--2024.

\bibitem[\protect\citeauthoryear{Goolam, Scialdone, Graham, MacAulay, and
  Jedrusik}{Goolam et~al.}{2016}]{goolam2016}
Goolam, M., A.~Scialdone, S.~J.~L. Graham, I.~C. MacAulay, and A.~Jedrusik
  (2016).
\newblock Heterogeneity in oct4 and sox2 targets biases cell fate in 4-cell
  mouse embryos.
\newblock {\em Cell\/}~{\em 165}, 61--74.

\bibitem[\protect\citeauthoryear{Holland, Laskey, and Leinhardt}{Holland
  et~al.}{1983}]{holland83}
Holland, P.~W., K.~B. Laskey, and S.~Leinhardt (1983).
\newblock Stochastic block models: First steps.
\newblock {\em Social Networks\/}~{\em 5}, 109--137.

\bibitem[\protect\citeauthoryear{Hou, Ji, Ji, and Hicks}{Hou
  et~al.}{2020}]{sparseScRNA}
Hou, W., Z.~Ji, H.~Ji, and S.~Hicks (2020).
\newblock A systematic evaluation of single-cell rna-sequencing imputation
  methods.
\newblock {\em Genome Biology\/}~{\em 21\/}(218).

\bibitem[\protect\citeauthoryear{Joseph and Yu}{Joseph and
  Yu}{2016}]{joseph2016}
Joseph, A. and B.~Yu (2016).
\newblock Impact of regularization on spectral clustering.
\newblock {\em Annals of Statistics\/}~{\em 44\/}(4), 1765--1791.

\bibitem[\protect\citeauthoryear{Karrer and Newman}{Karrer and
  Newman}{2011}]{karrer2011}
Karrer, B. and E.~J. Newman (2011).
\newblock Stochastic blockmodels and community structure in networks.
\newblock {\em Physics Review E\/}~{\em 83\/}(1).

\bibitem[\protect\citeauthoryear{Kiselev, Andrews, and Hemberg}{Kiselev
  et~al.}{2019}]{kiselev2019}
Kiselev, V.~Y., T.~S. Andrews, and M.~Hemberg (2019).
\newblock Chalenges in unsupervised clustering of single-cell rna-seq data.
\newblock {\em Nature Reviews Genetics\/}~{\em 20}, 273--282.

\bibitem[\protect\citeauthoryear{Kiselev, Kirschner, Schaub, Andrews, Yiu,
  Chandra, Natarajan, Reik, Barahona, Green, and Hemberg}{Kiselev
  et~al.}{2017}]{SC3}
Kiselev, V.~Y., K.~Kirschner, M.~T. Schaub, T.~Andrews, A.~Yiu, T.~Chandra,
  K.~N. Natarajan, W.~Reik, M.~Barahona, A.~R. Green, and M.~Hemberg (2017).
\newblock Sc3: consensus clustering of single-cell rna-seq data.
\newblock {\em Nature Methods\/}~{\em 14}, 483--486.

\bibitem[\protect\citeauthoryear{Kiselev, Kristina, Schaub, and
  Andrews}{Kiselev et~al.}{2017}]{kiselev2017}
Kiselev, V.~Y., K.~Kristina, M.~T. Schaub, and T.~e.~a. Andrews (2017).
\newblock Sc3- consensus clustering of single-cell rna-seq data.
\newblock {\em Nature Methods\/}~{\em 14\/}(5), 483--486.

\bibitem[\protect\citeauthoryear{Kolodziejczyk, Kim, Tsang, Ilicic, and
  Henriksson}{Kolodziejczyk et~al.}{2016}]{kolodziejczyk2016}
Kolodziejczyk, A.~A., J.~K. Kim, J.~C.~H. Tsang, T.~Ilicic, and J.~Henriksson
  (2016).
\newblock Single cell rna-sequencing of pluripotent states unlocks modular
  transcriptional variation.
\newblock {\em Cell Stem Cell\/}~{\em 17}, 471--485.

\bibitem[\protect\citeauthoryear{Lancichinetti and Fortunato}{Lancichinetti and
  Fortunato}{2009}]{Lancichinetti2009}
Lancichinetti, A. and S.~Fortunato (2009).
\newblock Community detection algorithms : A comparative analysis.

\bibitem[\protect\citeauthoryear{Lee and Levina}{Lee and
  Levina}{2015}]{can2015}
Lee, C.~M. and E.~Levina (2015).
\newblock Estimating the true number of communities in networks by spectral
  methods.
\newblock {\em arXiv, https://arxiv.org/pdf/1507.00827.pdf\/}.

\bibitem[\protect\citeauthoryear{Lee and Schnelli}{Lee and
  Schnelli}{2018}]{lee2018}
Lee, J.~O. and K.~Schnelli (2018).
\newblock Local law and tracy-widom limit for sparse random matrices.
\newblock {\em Probability Theory and Related Fields\/}~{\em 171}, 543--616.

\bibitem[\protect\citeauthoryear{Lei}{Lei}{2016}]{lei16}
Lei, J. (2016).
\newblock A goodness-of-fit test for stochastic block models.
\newblock {\em The Annals of Statistics\/}~{\em 44\/}(1), 401--424.

\bibitem[\protect\citeauthoryear{Lei and Rinaldo}{Lei and
  Rinaldo}{2015}]{lei15}
Lei, J. and A.~Rinaldo (2015).
\newblock Consistency of spectral clustering in sparse stochastic block models.
\newblock {\em Annals of Statistics\/}~{\em 43\/}(1), 215--237.

\bibitem[\protect\citeauthoryear{Lei and Zhu}{Lei and Zhu}{}]{lei2014}
Lei, J. and L.~Zhu.
\newblock A generic sample splitting approach for refined community recovery in
  stochastic block models.
\newblock {\em Preprint. Available at arXiv:1411.1469\/}.

\bibitem[\protect\citeauthoryear{Li, Lei, Bhattacharya, Berge, Sarkar, Bickel,
  and Levina}{Li et~al.}{2020}]{li2020}
Li, T., L.~Lei, S.~Bhattacharya, K.~V.~d. Berge, P.~Sarkar, P.~J. Bickel, and
  E.~Levina (2020).
\newblock Hierarchical community detection by recursive partitioning.
\newblock {\em Journal of the American Statistical Association\/}.

\bibitem[\protect\citeauthoryear{Li, Levina, and Zhu}{Li et~al.}{2016}]{li2016}
Li, T., E.~Levina, and J.~Zhu (2016).
\newblock Network cross-validation by edge sampling.
\newblock {\em arXiv preprint arXiv:1612.04717\/}.

\bibitem[\protect\citeauthoryear{Luecken and Theis}{Luecken and
  Theis}{2019}]{best_practice_scRNA}
Luecken, M.~D. and F.~J. Theis (2019).
\newblock {\em Molecular Systems Biology\/}~{\em 15\/}(6).

\bibitem[\protect\citeauthoryear{Lyu}{Lyu}{2019}]{lyu2019}
Lyu, Y. e.~a. (2019).
\newblock {\em
  https://www.biorxiv.org/content/10.1101/768143v1.full;bioarXiv\/}.

\bibitem[\protect\citeauthoryear{Ma, Su, and Zhang}{Ma et~al.}{2019}]{ma2019}
Ma, S., L.~Su, and Y.~Zhang (2019).
\newblock Determining the number of communities in degree-corrected stochastic
  block models.
\newblock {\em arXiv:https://arxiv.org/pdf/1809.01028.pdf\/}.

\bibitem[\protect\citeauthoryear{Myers, von Lersner, Robbins, and Sang}{Myers
  et~al.}{2015}]{diffgenes2015}
Myers, J., A.~von Lersner, C.~Robbins, and Q.-X. Sang (2015).
\newblock Differentially expressed genes and signature pathways of human
  prostate cancer.
\newblock {\em PLoS One\/}, 10:e0145322.

\bibitem[\protect\citeauthoryear{Newman and Girvan}{Newman and
  Girvan}{2004}]{newman04}
Newman, M. E.~J. and M.~Girvan (2004).
\newblock Finding and evaluating community structures in networks.
\newblock {\em Physical Review E\/}~{\em 69}, 026113.

\bibitem[\protect\citeauthoryear{Peel and Clauset}{Peel and
  Clauset}{2015}]{peel2015}
Peel, L. and A.~Clauset (2015).
\newblock Detecting change points in the large-scale structure of evolving
  networks.
\newblock {\em AAAI\/}, 2914--2920.

\bibitem[\protect\citeauthoryear{Pollen, Nowaowski, Shuga, Wang, and
  Leyrat}{Pollen et~al.}{2014}]{pollen2014}
Pollen, A.~A., T.~J. Nowaowski, J.~Shuga, X.~Wang, and A.~A. Leyrat (2014).
\newblock Low-coverage single-cell mrna sequencing reveals cellular
  heterogeneity and activating signalling pathways in developing cerebral
  cortex.
\newblock {\em Nat Biotechnology\/}~{\em 32}, 1053--1058.

\bibitem[\protect\citeauthoryear{Qin and Rohe}{Qin and
  Rohe}{2013}]{qin_dcsbm2013}
Qin, T. and K.~Rohe (2013).
\newblock Regularized spectral clustering under the degree-corrected stochastic
  blockmodel.
\newblock {\em NIPS\/}.

\bibitem[\protect\citeauthoryear{Rohe, Chatterjee, and Yu}{Rohe
  et~al.}{2011}]{rohe11}
Rohe, K., S.~Chatterjee, and B.~Yu (2011).
\newblock Spectral clustering and the high-dimensional stochastic blockmodel.
\newblock {\em The Annals of Statistics\/}~{\em 39\/}(4), 1878--1915.

\bibitem[\protect\citeauthoryear{Satija, Farrell, Gennert, Schier, and
  Regev}{Satija et~al.}{2015}]{satija2015}
Satija, R., J.~A. Farrell, D.~Gennert, A.~F. Schier, and A.~Regev (2015).
\newblock Spatial reconstruction of single-cell gene expression data.
\newblock {\em Nature Biotechnology\/}~{\em 33\/}(5), 495--502.

\bibitem[\protect\citeauthoryear{Svensson, Beltrame, and Pachter}{Svensson
  et~al.}{2020}]{valentine2020}
Svensson, V., E.~d.~V. Beltrame, and L.~Pachter (2020).
\newblock A curated database reveals trends in single-cell transcriptomics.
\newblock {\em Database https://doi.org/10.1093/database/baaa073\/}.

\bibitem[\protect\citeauthoryear{Wang, Ramazzotti, Sano, Zhu, Pierson, and
  Batzoglou}{Wang et~al.}{2018}]{SIMLR}
Wang, B., D.~Ramazzotti, L.~D. Sano, J.~Zhu, E.~Pierson, and S.~Batzoglou
  (2018).
\newblock Simlr: A tool for large-scale genomic analyses by multi-kernel
  learning.
\newblock {\em Proteomics 10.1002/pmic.201700232\/}~{\em 18\/}(2).

\bibitem[\protect\citeauthoryear{Wang and Bickel}{Wang and
  Bickel}{2017}]{wang17}
Wang, R. Y.~X. and P.~J. Bickel (2017).
\newblock Likelihood-based model selection for stochastic block models.
\newblock {\em The Annals of Statistics\/}~{\em 45\/}(2), 500--528.

\bibitem[\protect\citeauthoryear{Yan, Yang, Guo, Yang, and Wu}{Yan
  et~al.}{2013}]{yan2013}
Yan, L., M.~Yang, H.~Guo, L.~Yang, and J.~Wu (2013).
\newblock Single-cell rna-seq profiling of human preimplantation embryos and
  embryonic stem cells.
\newblock {\em Nat Struct Mol Biol\/}~{\em 20}, 1131--1139.

\bibitem[\protect\citeauthoryear{Zhao}{Zhao}{2017}]{zhao17}
Zhao, Y. (2017).
\newblock A survey on theoretical advances of community detection in networks.
\newblock {\em WIREs Comput Stat\/}~{\em 9}, c1403.

\bibitem[\protect\citeauthoryear{Zhao, Levina, and Zhu}{Zhao
  et~al.}{2012}]{zhao12}
Zhao, Y., E.~Levina, and J.~Zhu (2012).
\newblock Consistency of community detection in networks under degree-corrected
  stochastic block models.
\newblock {\em The Annals of Statistics\/}~{\em 40\/}(4), 2266--2292.

\end{thebibliography}

\end{document}